\documentclass{PoS}
\usepackage{amsmath}
\usepackage{graphicx}

\title{The charm-quark contribution to $\epsilon_K$ and $\Delta M_K$}

\ShortTitle{The charm-quark contribution to $\epsilon_K$ and $\Delta M_K$}

\author{\speaker{Joachim Brod}%
  \\
  Excellence Cluster Universe, Technische Universit\"at M\"unchen,
  Garching, Germany\\
  University of Cincinnati, Cincinnati, OH, USA\\
  E-mail: \email{joachim.brod@uc.edu}}

\abstract{Neutral Kaon mixing plays an important role in the
  phenomenology of the standard model and its extensions because of
  its sensitivity to high-energy scales. In particular $\epsilon_K$,
  parameterising indirect $CP$ violation in the neutral Kaon system,
  serves as an important constraint on models of new physics and is
  well suited for the indirect search for heavy new particles. In
  order to exploit this potential, a precise prediction of the
  standard-model background is crucial. I give a short summary of the
  standard-model prediction of $\epsilon_K$, and present our recent
  NNLO QCD calculation of the charm-quark contribution $\eta_{cc}$ to
  the $|\Delta S| = 2$ effective Hamiltonian. We find a large $36\%$
  shift with respect to the NLO value that leads to
  $\eta_{cc}=1.87(76)$, shifting the standard-model prediction to
  $|\epsilon_K| = 1.81(28)\times 10^{-3}$.}

\FullConference{XXIst International Europhysics Conference on High-Energy Physics\\
		 21-27 July 2011\\
		 Grenoble, Rh\^ones-Alpes, France}

\begin{document}

\section{Introduction}

\begin{figure}[t]
\centering
\includegraphics[width=0.3\textwidth]{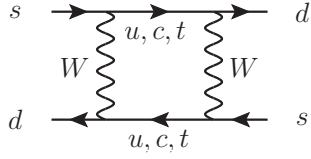}
\caption{LO box diagrams inducing the $|\Delta
  S|=2$ transition in the SM. \label{fig:eK}}
\end{figure}

Neutral Kaon mixing proceeds via the quark-level flavour-changing
neutral current (FCNC) $s-d$ transition. In the standard model of
particle physics (SM) it is forbidden at tree level and induced by the
weak interaction via the well-known box diagrams (see
Fig.~\ref{fig:eK}). The parameter $\epsilon_K$ describes indirect $CP$
violation in the neutral Kaon system. The top-quark contribution to
$\epsilon_K$ is parametrically suppressed by small CKM-matrix
elements, whereas the hadronic matrix elements of non-SM operators are
enhanced by QCD effects. This leads to an exceptional sensitivity to
high-energy scales.

In the last decade there has been huge progress in reducing the theory
uncertainty of the hadronic contribution to $\epsilon_K$, in
particular in the lattice computation of the hadronic matrix elements.
We have in turn calculated the NNLO QCD corrections to the effective
$|\Delta S|=2$ Hamiltonian. In this talk, I present our recent NNLO
calculation of the charm-quark contribution $\eta_{cc}$: the
correction is large and shifts the SM prediction to $|\epsilon_K| =
1.81(28)\times 10^{-3}$~\cite{Brod:2011ty}. This prediction is
in slight disagreement with the precisely measured value
$|\epsilon_K|=2.228(11)\times 10^{-3}$~\cite{PDG2010}. However,
the remaining sizeable uncertainty of the SM prediction currently
precludes us from inferring a clear sign of new physics in neutral
Kaon mixing.

\boldmath
\section{SM prediction of $\epsilon_K$}\label{sec:sd}
\unboldmath

The parameter $\epsilon_K$ is defined as the ratio of decay
amplitudes\footnote{A very lucid discussion of the details can be
  found in Ref.~\cite{Anikeev:2001rk}.}  $ \epsilon_K = \langle (\pi
\pi)_{I=0} \lvert K_L \rangle / \langle (\pi \pi)_{I=0} \lvert K_S
\rangle$ and vanishes in case of exact $CP$ symmetry. It can be
expressed by the following phenomenological
formula~\cite{Anikeev:2001rk}
\begin{equation}\label{eq:eK}
\epsilon_K = e^{i\phi_\epsilon} \sin
\phi_\epsilon \left( \frac{\text{Im}\, M_{12}}{\Delta M_K} + \xi
\right)\, ,
\end{equation}
where $\phi_\epsilon = \arctan(2\Delta M_K/\Delta\Gamma_K) \approx
45^\circ$. $\xi=\text{Im}\,A_0/\text{Re}\,A_0$, where $A_I=\langle
(\pi \pi)_{I} \lvert K^0 \rangle$, is a non-perturbative correction of
$\mathcal{O}(5\%)$. We have neglected terms proportional to the second
power of the small quantities $\phi=\arg(-M_{12}/\Gamma_{12})$ and
$|A_2|/|A_0|$, and to the first power in $\Gamma_L/\Gamma_S$. Taking
$\Delta M_K$ and $\phi_\epsilon$ from experiment, we obtain a theory
prediction of $\epsilon_K$ by computing $\text{Im}\, M_{12}$ and
$\xi$.

The effective Hamiltonian below the charm-quark scale, inducing the
$|\Delta S|=2$ transitions in the SM, is given by
\begin{equation}
    \mathcal{H}^{\Delta S=2}_{} = \frac{G_F^2}{4 \pi^2} M_W^2 \left[
      \lambda_c^2 \eta_{cc} S(x_c) +
      \lambda_t^2 \eta_{tt} S(x_t) +
      2 \lambda_c \lambda_t \eta_{ct} S(x_c,x_t) \right] b(\mu)
    \tilde Q_{S2} + \textrm{H.c.} + \dots \, ,
  \label{eq:Hlo}
\end{equation}
where $G_F$ is the Fermi constant, $M_W$ the $W$-boson mass, and $x_i
= m_i^2/M_W^2$ ($m_i$ being the quark masses). We have defined
$\lambda_i = V_{is}^* V_{id}$ and eliminated $\lambda_u$ using the
unitarity of the CKM matrix. The factor $b(\mu)$ contains the
remaining scale dependence. $M_{12}$ is then given in terms of the
effective Hamiltonian by $M_{12} = \langle \bar K^0
|\mathcal{H}^{|\Delta S|=2}| K^0 \rangle / (2 M_K)$.

In Eq.~\eqref{eq:Hlo}, $\tilde Q_{S2} = (\overline{s}_L \gamma_{\mu}
d_L)^2$ is the leading local four-quark operator that induces the
$|\Delta S|=2$ transition, defined in terms of the left-handed $s$-
and $d$-quark fields. Higher-dimensional operators are estimated to
contribute less than 1\% to
$\epsilon_K$~\cite{Cata:2003mn}. The hadronic matrix
elements of the operator $\tilde Q_{S2}$ constitute the major part of
the long-distance contributions to $\epsilon_K$. They are
parameterised by the bag factor
\begin{equation}
  \label{eq:bkpar}
  \hat B_K = \frac{3}{2} b(\mu) 
  \frac{
    \langle \bar K^0 | \tilde Q_{S2} | K^0\rangle}{
    f_K^2 M_K^2}\, , 
\end{equation}
where $f_K$ is the Kaon decay constant, and $b(\mu)$ is factored out
of Eq.~\eqref{eq:Hlo} in such a way that $\hat B_K$ is a
renormalisation-group invariant quantity. It has been calculated
precisely using lattice QCD, with an total uncertainty of $4\%$ or
less~\cite{Aubin:2009jh}. The obtained
values are consistent with the upper bound derived in the framework of
large-$N$ QCD~\cite{Gerard:2010jt}.

There are additional long-distance contributions which are not
contained in $\hat B_K$, proportional to the dispersive and absorptive
parts of the amplitude $\int d^4x \langle \bar K^0 |
\mathcal{H}^{\Delta S=1}(x) \mathcal{H}^{\Delta S=1}(0) | K^0
\rangle$, respectively. The estimates of the parameter
$\xi$~\cite{Anikeev:2001rk,Buras:2008nn} and of the dispersive part of
the amplitude contributing to $\text{Im}\,
M_{12}$~\cite{Buras:2010pza} have been combined with the experimental
value of $\phi_\epsilon$ into the correction factor
$\kappa_\epsilon=0.94(2)$~\cite{Buras:2010pza}, which multiplies the
expression~\eqref{eq:eK}, in which we then set $\xi=0$ and
$\phi_\epsilon=45^\circ$.

The short-distance contributions are contained in the loop functions
$S$ in Eq.~\eqref{eq:Hlo}, the $\eta$ factors comprising the
higher-order QCD corrections. The dominant term proportional to
$\lambda_t^2$ contributes approximately $+75\%$ to $\epsilon_K$. The
QCD corrections have been computed by a fixed-order matching
calculation at the top-quark scale at NLO, yielding
$\eta_{tt}=0.5765(65)$~\cite{Buras:1990fn}.  The term proportional to
$\lambda_c \lambda_t$ contributes roughly $+43\%$ to $\epsilon_K$. Our
NNLO QCD calculation~\cite{Brod:2010mj} leads to
$\eta_{ct}=0.496(47)$. The smallest contribution of about $-18\%$ is
proportional to $\lambda_c^2$. The GIM mechanism causes the absence of
a large logarithm $\log(m_c/M_W)$ at LO and the mixing of $|\Delta
S|=1$ into $|\Delta S|=2$ operators above the charm-quark scale. We
performed a three-loop matching calculation at the charm-quark
scale~\cite{Brod:2011ty} and find $\eta_{cc}=1.87(76)$. The large
$+36\%$ shift with respect to the NLO result~\cite{Herrlich:1996vf}
and the large residual scale dependence at NNLO give rise to our
assignment of a sizeable theory uncertainty. A conversion of our
result to a suitable RI-SMOM renormalisation
scheme~\cite{Sturm:2009kb} might improve the convergence of the
perturbation series.

Using the numerical input values as in~\cite{Brod:2011ty} and the
NNLO values $\eta_{ct}=0.496(47)$ and $\eta_{cc}=1.87(76)$, we obtain
the SM prediction
\begin{equation}
  |\epsilon_K| = (1.81\pm 0.14_{\eta_{cc}} \pm 0.02_{\eta_{tt}}
  \pm 0.07_{\eta_{ct}} \pm 0.08_{\text{LD}} \pm 0.23_{\text{parametric}})
  \times 10^{-3} \, .
\end{equation}

The error indicated by LD originates from the long-distance
contributions, namely $\hat B_K$ and $\kappa_\epsilon$. A large part
of the parametric error stems from $|V_{cb}|$, which enters the
dominant top-quark contribution to the fourth power.

Finally, we remark in passing that using the NNLO value of $\eta_{cc}$
and again the input as in Ref.~\cite{Brod:2011ty}, the short-distance
contribution to $\Delta M_K$ accounts for $86(34)\%$ of the
measured value~\cite{PDG2010}.

\section{Conclusion and outlook}

I have presented the NNLO SM prediction of the important observable
$\epsilon_K$, with the result $|\epsilon_k| = 1.81(28)\times
10^{-3}$. The uncertainty is dominated by the error on $|V_{cb}|$ and
the theory uncertainty of the perturbative contribution. The latter
can possibly be decreased by transforming the result to a suitable
RI-SMOM scheme at NNLO, or, in the long run, by computing the effects
of a dynamical charm quark using lattice QCD.

\begin{acknowledgments}
I thank M.~Gorbahn for the fruitful and pleasant collaboration and the
organisers of EPS~2011 for the interesting and inspiring conference.
\end{acknowledgments}


\end{document}